\documentstyle[prd,aps,epsfig]{revtex}

\def\be{\begin{equation}}
\def\ee{\end{equation}}
\def\ba{\begin{eqnarray}}
\def\ea{\end{eqnarray}}

\begin{document}  
\draft

\title{Diagonalization of Quark Mass Matrices
and the Cabibbo-Kobayashi-Maskawa Matrix}

\author{Andrija Ra\v{s}in}

\address{{\it
International Center for Theoretical
Physics, Strada Costiera 11, 34100 Trieste, Italy }}

\maketitle

.	

\begin{abstract}

I discuss some general aspects of diagonalizing the quark mass matrices 
and list all possible parametrizations of the 
Cabibbo-Kobayashi-Maskawa matrix (CKM) in terms of
three rotation angles and a phase. I systematically study the relation
between the rotations needed to diagonalize the Yukawa matrices
and various parametrizations of the CKM.

\end{abstract}
\pacs{hep-ph/9708216}

\vspace{0.2in}

{\bf A. Introduction}\hspace{0.5cm} 

The most general Cabibbo-Kobayashi-Maskawa matrix (CKM)
\cite{cabi63,koba73} can be written as a function of three angles and one
phase. Various parametrizations of CKM exist
today\cite{koba73,maia76,maia77,chau84,pdgr96,dimo92}
in which these three angles and one phase appear in various places.
These four physical parameters along with the six quark masses
are derived from the up and down quark mass matrices, which
are in general arbitrary complex matrices with 9 real parameters
and 9 imaginary parameters each. 

Usually general arguments are used to do the counting of parameters and
some ``trial and error" to set up a particular parametrization for the CKM
matrix. However, to a model builder who predicts a particular pattern for
the Yukawa matrices it may prove useful to see how exactly one obtains
the parametrization from the original quark mass matrices. In the
literature there exist many examples of arguing in favor of a
particular parametrization of the CKM for some pattern of quark mass 
matrices\cite{dimo92,barb96,frit97}.
However, a systematic study of an arbitrary case is still lacking.

In this paper I lay out a procedure of diagonalizing the quark
mass matrices and I numerate all possible parametrizations
of the CKM matrix in terms of three rotation angles and one phase. 
I first start in Section B with some preliminaries about diagonalizing
arbitrary complex matrices and offer another way of counting the
number of parameters. In Section C I explicitly parametrize arbitrary
$2\times2$ and $3\times3$ unitary matrices in terms of 
rotation angles and phases. The CKM parametrizations follow trivially
by a phase redefinition and I list all possible parametrizations in a
Table. Next, I turn to explicit relations between the parameters of the
CKM and the elements of the quark mass matrices. The case of two
generations is studied in Sections D (diagonalization) and E (CKM). This
case serves as a proof of some simple results that become useful in
studying the case of three generations in Sections F and G. 
I show how different parametrizations of the CKM come about
and how some of them are more suitable for some quark mass matrix 
structures.

As inspired by the difference in the quark masses, I will assume in
this paper that all eigenvalues of the mass matrices are different,
although the discussion is easily generalized to the case of degenerate
masses. Also the discussion easily generalizes to an arbitrary number of
generations. 

\vspace{0.2in}

{\bf B. Generalities on the diagonalization of an arbitrary
complex square matrix}\hspace{0.5cm}

An arbitrary $n\times n$ complex matrix ${\bf h}$ is diagonalized by two
unitary
matrices ${\bf U}$ and ${\bf V}$
\be
{\bf m} = {\bf U}^\dagger {\bf h} {\bf V}
\label{first}
\ee
where ${\bf m}$ is a diagonal matrix with real and nonnegative entries on
the diagonal. The matrices
${\bf U}$ and ${\bf V}$  diagonalize the following products of ${\bf h}$
\be
{\bf m^2} = {\bf U}^\dagger {\bf h} {\bf h}^\dagger {\bf U} \, \, , \, \,
{\bf m^2} = {\bf V}^\dagger {\bf h}^\dagger {\bf h} {\bf V} 
\label{defineuv}
\ee

It is known that an arbitrary unitary matrix has $n(n-1)/2$ angles
and $n(n+1)/2$ phases.  Let us see how this counting works for 
${\bf U}$ and ${\bf V}$ and how many of these parameters
are fixed by matrix ${\bf h}$, the matrix
which we are diagonalizing, and how many are left completely arbitrary.
For example, let us look at the product ${\bf h}^\dagger {\bf h}$. It has
$n(n+1)/2$ angles and $n(n-1)/2$ phases
\footnote{In this article I will use the terms ``angles"
and ``phases" instead of real and imaginary parameters,
respectively.}.  Of the $n(n+1)/2$ angles,
$n$ describe the $n$ diagonal entries of ${\bf m}^2$ so that $n(n-1)/2$
angles must enter ${\bf V}$. Similarly, all the $n(n-1)/2$ phases must
enter ${\bf V}$, since ${\bf m}$ is real. Notice however that ${\bf V}$ is
not completely determined by (\ref{defineuv}). It can be multiplied from
the right by an arbitrary phase rotation which contains $n$ phases that
do not depend on ${\bf h}$
\be
{\bf V} \to {\bf V}
\left(
\begin{array}{ccc}
e^{i\alpha_1} & &\\
& ... & \\
& & e^{i\alpha_n}
\end{array}
\right)
\ee
giving a total of $n(n+1)/2$ phases for ${\bf V}$.
Similar counting is valid for ${\bf U}$. 

Finally, inverting the first
relation (\ref{first})
\be
{\bf h} = {\bf U} {\bf m} {\bf V}^\dagger
\label{inverted}
\ee
we can count the number of phases and angles on both sides. On the
lefthand side (lhs) we have $n^2$ angles and $n^2$ phases. This should be
matched by the righthand side (rhs). Let us count first the number
of angles. The matrices ${\bf U}$ and ${\bf V}$ have
each $n(n-1)/2$ angles together with $n$ real parameters in ${\bf m}$
give $2 (n(n-1)/2) + n = n^2$ which checks.
For the phases, ${\bf U}$ and ${\bf V}$ have
$n(n+1)/2$ each, of which $n(n-1)/2$ in each matrix is already fixed 
by the elements of ${\bf h}$ from
(\ref{defineuv}). But then notice that in (\ref{inverted}) of the
remaining pair of arbitrary $n$ phases in  ${\bf U}$ and ${\bf V}$ only
the difference of the phases appears and gets fixed by elements of ${\bf
h}$. Thus the number of phases on the rhs is $2 (n(n-1)/2) + n = n^2$
which checks again.

Thus, we conclude that ${\bf U}$ and ${\bf V}$ are unitary matrices with
$n(n-1)/2$ angles $n(n+1)/2$ phases each, of which there are altogether
$n$ phases left that are {\it not} fixed by ${\bf h}$. These $n$ arbitrary
phases can be chosen only among the phases that sit on the far right in a
product of matrices representing ${\bf U}$ or ${\bf V}$.

Let us now turn to the counting of parameters in CKM. 
Quark masses come from the following 
terms in the Lagrangian
\be
u^c {\bf h}^u Q  +  
d^c {\bf h}^d Q  
\ee
Each of the matrices ${\bf h}^{u,d}$ is diagonalized by the procedure
outlined above

\be
{\bf h}^{u,d} = {\bf U}^{u,d} {\bf m} {\bf V}^{u,d \, \dagger} 
\ee
so that the CKM matrix is given by
\be
{\bf K} = {\bf V}^{u\dagger} {\bf V}^d 
\ee

${\bf K}$ is a unitary matrix and can also in principle have $n(n-1)/2$
angles and $n(n+1)/2$ phases. However, some of the phases can be
rotated away. From the discussion above it follows that we can choose $n$
arbitrary phases in ${\bf V}^u$ and $n$ arbitrary phases in ${\bf V}^d$.
These phases sit on the far right of each of ${\bf V}^i$ so
that ${\bf K}$ has $n$ arbitrary phases on the left and $n$ arbitrary
phases on the right. Now all of these $2n$ phases can be used to redefine
phases in ${\bf K}$, except one. It is the phase proportional to unity
that must be set to zero\footnote{This freedom corresponds to the familiar
baryon number of the standard model.}. Thus
the CKM has $n(n+1)/2 - 2n + 1 = (n-1)(n-2)/2$ phases.   

\vspace{0.2in}

{\bf C. General form of a unitary matrix}\hspace{0.5cm} 

In this Section, I will write out the general forms of
$n \times n$ unitary matrices for $n=2,3$ with the obvious
generalization to the case of arbitrary $n$.
They follow from the definition of a
unitary  $n\times n$ matrix ${\bf V}$ 
\be
{\bf V} {\bf V}^\dagger = 
{\bf V}^\dagger {\bf V} = {\bf 1}
\label{unitcond}
\ee

{\bf General $2\times2$ unitary matrix}

Equations (\ref{unitcond}) can be written out in terms of the 
elements of ${\bf V}$
\ba
|V_{11}|^2 + |V_{12}|^2 = 1 
\label{u21}\\
V_{11} V_{21}^* + V_{12} V_{22}^* = 0 
\label{u22}\\
V_{21} V_{11}^* + V_{22} V_{12}^* = 0 
\label{u23}\\
|V_{21}|^2 + |V_{22}|^2 = 1 
\label{u24}\\
|V_{11}|^2 + |V_{21}|^2 = 1 
\label{u25}\\
V_{11}^* V_{12} + V_{21}^* V_{22} = 0
\label{u26}\\
V_{12}^* V_{11} + V_{22}^* V_{21} = 0 
\label{u27}\\
|V_{12}|^2 + |V_{22}|^2 = 1 
\label{u28}
\ea
From (\ref{u21}) we can introduce an angle $\theta$ so that
$|V_{11}| = \cos \theta \equiv c$ and
$|V_{12}| = \sin \theta \equiv s$. 
From (\ref{u24})-(\ref{u25}) also 
$|V_{21}| = s$ and
$|V_{22}| = c$. So far ${\bf V}$ is of the form
\be
{\bf V} =
\left(
\begin{array}{cc}
c e^{i\alpha_{11}} & s e^{i\alpha_{12}} \\
s e^{i\alpha_{21}} & c e^{i\alpha_{22}}
\end{array}
\right)
\ee
and we have to determine the phases from the remaining equations.
For example from (\ref{u22})
\be
c s e^{i(\alpha_{11}-\alpha_{21})} + s c 
e^{i(\alpha_{12}-\alpha_{22})} = 0
\ee
The other equations will not introduce any new constraint.
We can eliminate one phase, for example $\alpha_{21}$, so that
the most general form for an arbitrary unitary matrix is
\be
{\bf V} =
\left(
\begin{array}{cc}
c e^{i\alpha_{11}} & s e^{i\alpha_{12}} \\
- s e^{i(\alpha_{11}-\alpha_{12}+\alpha_{22})} & c e^{i\alpha_{22}}
\end{array}
\right)
=
\left(
\begin{array}{cc}
1 & \\
& e^{i(\alpha_{22}-\alpha_{12})} 
\end{array}
\right)
\left(
\begin{array}{cc}
c & s\\
-s & c 
\end{array}
\right)
\left(
\begin{array}{cc}
e^{i\alpha_{11}} & \\
& e^{i\alpha_{12}}
\end{array}
\right)
\ee
Thus, we obtain the familiar result that an arbitrary $2\times2$ unitary
matrix is described by one angle and three phases.
Notice the ambiguity in placing the three phases. They could have been
in any of the four diagonal positions in the left and right matrices 
on the rhs.

{\bf General $3\times3$ unitary matrix}

In the case of $n=3$ we see from (\ref{unitcond}) that the elements of a
unitary matrix ${\bf V}$ satisfy
\ba
|V_{11}|^2 + |V_{12}|^2 + |V_{13}|^2 = 1 
\label{ud311}\\
V_{11} V_{21}^* + V_{12} V_{22}^* + V_{13} V_{23}^* = 0
\label{ud312}\\
V_{11} V_{31}^* + V_{12} V_{32}^* + V_{13} V_{33}^* = 0
\label{ud313}\\
V_{21} V_{11}^* + V_{22} V_{12}^* + V_{23} V_{13}^* = 0
\label{ud321}\\
|V_{21}|^2 + |V_{22}|^2 + |V_{23}|^2 = 1 
\label{ud322}\\
V_{21} V_{31}^* + V_{22} V_{32}^* + V_{23} V_{33}^* = 0
\label{ud323}\\
V_{31} V_{11}^* + V_{32} V_{12}^* + V_{33} V_{13}^* = 0
\label{ud331}\\
V_{31} V_{21}^* + V_{32} V_{22}^* + V_{33} V_{23}^* = 0
\label{ud332}\\
|V_{31}|^2 + |V_{32}|^2 + |V_{33}|^2 = 1 
\label{ud333}\\
|V_{11}|^2 + |V_{21}|^2 + |V_{31}|^2 = 1 
\label{u311}\\
V_{11}^* V_{12} + V_{21}^* V_{22} + V_{31}^* V_{32} = 0
\label{u312}\\
V_{11}^* V_{13} + V_{21}^* V_{23} + V_{31}^* V_{33} = 0 
\label{u313}\\
V_{12}^* V_{11} + V_{22}^* V_{21} + V_{32}^* V_{31} = 0 
\label{u321}\\
|V_{12}|^2 + |V_{22}|^2 + |V_{32}|^2 = 1 
\label{u322}\\
V_{12}^* V_{13} + V_{22}^* V_{23} + V_{32}^* V_{33} = 0 
\label{u323}\\
V_{13}^* V_{11} + V_{23}^* V_{21} + V_{33}^* V_{31} = 0 
\label{u331}\\
V_{13}^* V_{12} + V_{23}^* V_{22} + V_{33}^* V_{32} = 0 
\label{u332}\\
|V_{13}|^2 + |V_{23}|^2 + |V_{33}|^2 = 1 
\label{u333}
\ea
I now proceed similarly to case $n=2$. I will first 
parametrize the ``diagonal" equations 
(\ref{ud311}),(\ref{ud322}),(\ref{ud333}),
(\ref{u311}),(\ref{u322}) and (\ref{u333}). 
This is the place where various parametrizations
come in. Let us first do the ``standard"
parametrization\cite{maia77,chau84}. We
introduce the first angle $\theta_{13}$ such that 
\footnote{As will be obvious later, the choice of the two indices on the
angle represent the plane in which a rotation is made. One might think
of using a nicer notation with one index only, (i.e. $\theta_2$) to denote
the axis of rotation, but this will make less transparent the meaning of
the angles and not possible to generalize to more generations. And it 
would certainly draw criticism ($\theta_3$ would be the Cabibbo
angle?!).} 
\be
|V_{13}| = \sin \theta_{13} \equiv s_{13}
\ee
The various parametrizations will introduce this first angle at
a different element (see the Table later).
It follows from (\ref{ud311}) and (\ref{u333}) that  
\be
|V_{11}|^2 + |V_{12}|^2 = c^2_{13} \, , \,
|V_{23}|^2 + |V_{33}|^2 = c^2_{13} 
\ee
This motivates us to introduce two more angles $\theta_{12}$
and $\theta_{23}$ such that
\ba
|V_{11}|^2 = c^2_{12} c^2_{13} \, , \,
& |V_{12}|^2 = s^2_{12} c^2_{13}  \nonumber \\
|V_{23}|^2 = s^2_{23} c^2_{13} \, , \,
& |V_{33}|^2 = c^2_{23} c^2_{13}
\ea
This way of introducing additional angles is the only one that allows
the interpretation of these angles as rotation angles\footnote{An 
alternative choice would be that we choose another element of ${\bf V}$
to be a single angle, for example 
$V_{12} = s_{12}$ \cite{king95}. However,
I view this as unattractive since then 
$V_{11}$ is forced to be a 
$\sqrt{c^2_{13}-s^2_{12}}$ and it certainly does not allow an
interpretation as a rotation angle, so that $s_{12}$ is just doubling of
the name for $V_{12}$.}.

Since of the 6 diagonal equations only 4 are independent,
it is enough to introduce just one more parameter, to take care
of all the diagonal equations. I choose a phase $\delta$
such that
\ba
|V_{21}|^2 = |s_{12} c_{23} + c_{12} s_{23} s_{13} e^{i\delta}|^2 \,
, \,
|V_{22}|^2 = |c_{12} c_{23} - s_{12} s_{23} s_{13} e^{i\delta}|^2
\nonumber\\
|V_{31}|^2 = |s_{12} s_{23} - c_{12} c_{23} s_{13} e^{i\delta}|^2 \,
, \,
|V_{32}|^2 = |c_{12} s_{23} + s_{12} c_{23} s_{13} e^{i\delta}|^2
\ea

Now we are ready to look at the off-diagonal equations, and determine
the phases of elements of ${\bf V}$. Let us define the yet unknown
phases as $V_{11} = c_{12}c_{13}e^{i\alpha_{11}}$, 
$V_{21} = (s_{12} c_{23} + c_{12} s_{23} s_{13}
e^{i\delta}) e^{i\alpha_{21}}$, {\it etc}, and
$V_{13}=s_{13}e^{i(\alpha_{13}-\delta)}$.
From (\ref{ud312}) we have
\ba
& 
c_{12} c_{13} (s_{12} c_{23} + c_{12} s_{23} s_{13} e^{-i\delta})
e^{i(\alpha_{11}-\alpha_{21})} \nonumber\\
+ & 
s_{12} c_{13} (c_{12} c_{23} - s_{12} s_{23} s_{13} e^{-i\delta})
e^{i(\alpha_{12}-\alpha_{22})} \nonumber\\
+ &
s_{13} s_{23} c_{13} e^{-i\delta} e^{i(\alpha_{13}-\alpha_{23})} = 0
\ea
This should be valid for any 
$\theta_{12}$, $\theta_{13}$, $\theta_{23}$ 
and $\delta$. For $\theta_{12}=0$ or for $\theta_{13}=0$ we
get respectively
\be
e^{i(\alpha_{11}-\alpha_{21})} = - e^{i(\alpha_{13}-\alpha_{23})} \, , \,
e^{i(\alpha_{11}-\alpha_{21})} = - e^{i(\alpha_{12}-\alpha_{22})}
\ee
Similarly, from (\ref{ud313})
\be
e^{i(\alpha_{11}-\alpha_{31})} = e^{i(\alpha_{13}-\alpha_{33})} \, , \,
e^{i(\alpha_{11}-\alpha_{31})} = - e^{i(\alpha_{12}-\alpha_{32})}
\ee
All other equations are not independent and introduce no new constraints.
Thus, out of the 9 $\alpha_{ij}$, there will be 5 independent ones. We
choose $\alpha_{11}$, $\alpha_{12}$, $\alpha_{13}$, $\alpha_{23}$
and $\alpha_{33}$. Then the most general unitary matrix ${\bf V}$ is
given by
\be 
{\bf V} = 
\left( 
\begin{array}{ccc}
1 & & \\
& e^{i(\alpha_{23}-\alpha_{13})}  & \\
& & e^{i(\alpha_{33}-\alpha_{13})} \\
\end{array}
\right)
\left(
\begin{array}{ccc}
c_{12} c_{13} &
s_{12} c_{13} &
s_{13} e^{-i\delta}\\
-s_{12} c_{23} - c_{12} s_{23} s_{13} e^{i\delta}  &
c_{12} c_{23} - s_{12} s_{23} s_{13} e^{i\delta}  &
s_{23} c_{13} \\
s_{12} s_{23} - c_{12} c_{23} s_{13} e^{i\delta} &
-c_{12} s_{23} - s_{12} c_{23} s_{13} e^{i\delta}  &
c_{23} c_{13} 
\end{array}
\right)
\left( 
\begin{array}{ccc}
e^{i\alpha_{11}} & & \\
& e^{i\alpha_{12}}  & \\
& & e^{i\alpha_{13}} \\
\end{array}
\right)
\label{firstunitary}
\ee
Notice that ${\bf V}$ can be broken up into a product of matrices
\ba
{\bf V} & = &
\left( 
\begin{array}{ccc}
1 & & \\
& e^{i(\alpha_{23}-\alpha_{13})}  & \\
& & e^{i(\alpha_{33}-\alpha_{13})} \\
\end{array}
\right)
\nonumber\\
& \times &
\left(
\begin{array}{ccc}
1 & & \\
& c_{23} & s_{23} \\
& -s_{23} & c_{23}
\end{array}
\right)
\left( 
\begin{array}{ccc}
e^{-i\delta} & & \\
& 1  & \\
& & 1 \\
\end{array}
\right)
\left( 
\begin{array}{ccc}
c_{13} & & s_{13} \\
& 1 &  \\
-s_{13} &  & c_{13}
\end{array}
\right)
\left( 
\begin{array}{ccc}
e^{i\delta} & & \\
& 1  & \\
& & 1 \\
\end{array}
\right)
\left( 
\begin{array}{ccc}
c_{12} & s_{12} & \\
-s_{12} & c_{12} & \\
& & 1
\end{array}
\right)
\nonumber\\
& \times &
\left( 
\begin{array}{ccc}
e^{i\alpha_{11}} & & \\
& e^{i\alpha_{12}}  & \\
& & e^{i\alpha_{13}} \\
\end{array}
\right)
\ea
or in an obvious symbolic notation 
\be
{\bf V} = (1 \alpha_4 \alpha_5 ) 
{\bf R}_{23} (-\delta 1 1) {\bf
R}_{13} (\delta 1 1) {\bf R}_{12} 
(\alpha_1 \alpha_2 \alpha_3)
\label{genunit}
\ee
where 
\be
(\alpha_1 \alpha_2 \alpha_3) = diag (e^{i\alpha_1},
e^{i\alpha_2},e^{i\alpha_3}) \, \, , \, \, 
(1 \alpha_4 \alpha_5) = diag (1, e^{i\alpha_4},e^{i\alpha_5}) 
\ee
and I renamed $\alpha$s: $\alpha_1=\alpha_{11}$,
$\alpha_2=\alpha_{12}$, {\it etc}.
${\bf V}$ has 3 angles and 6 phases, as it should be for an
arbitrary $3\times3$ unitary matrix. Notice again the ambiguity in placing
the 5 $\alpha$s among the six diagonal entries of the far left and far
right matrices on the rhs. Also notice that the angle $\delta$ could have
been placed at different places in the middle matrix, by an appropriate
redefinition of $\alpha$s. However, what is important to notice is that
one combination of phases cannot be pulled out between 
the three rotations.

An important note is in order here regarding the 
Cabibbo-Kobayashi-Maskawa matrix. The CKM
matrix is an arbitrary unitary matrix with five phases rotated away
through the phase redefinition of the left handed up and down quark
fields.  Notice that the middle part of the rhs of (\ref{firstunitary}) is
exactly the ``standard" parametrization\cite{maia77,chau84,pdgr96}
\be
{\bf K} = 
{\bf R}_{23} (-\delta 1 1) {\bf R}_{13} (\delta 1 1) {\bf R}_{12} 
\ee
As we saw in the previous section, it is exactly the freedom of the
far right phases in ${\bf V^u}$ and ${\bf V^d}$ that allows us to
eliminate all the phases $\alpha_i$, but not the phase $\delta$.

Now that we have one parametrization explicitly worked out it is
easy to obtain any other parametrization. One just starts by picking
one element of the unitary matrix ${\bf V}$ to be a single sine or
a cosine of an angle, and proceeds by first looking at the diagonal
equations to parametrize the other two angles and the phase $\delta$.
Then, from the off diagonal equations,
one fixes all the other phases in terms of five phases.
I list all possibilities in Table I. 
\begin{table}
\be
\begin{array}{c|c|c}
& {\bf K} &  J \\
\hline
1 & {\bf R}_{23}(-\delta11){\bf R}_{13}(\delta11){\bf R}_{12} =
\left(
\begin{array}{ccc}
c_{12} c_{13} &
s_{12} c_{13} &
s_{13} e^{-i\delta}\\
-s_{12} c_{23} - c_{12} s_{23} s_{13} e^{i\delta}  &
c_{12} c_{23} - s_{12} s_{23} s_{13} e^{i\delta}  &
s_{23} c_{13} \\
s_{12} s_{23} - c_{12} c_{23} s_{13} e^{i\delta} &
-c_{12} s_{23} - s_{12} c_{23} s_{13} e^{i\delta}  &
c_{23} c_{13} 
\end{array}
\right)
& 
s_{12}s_{13}s_{23}c_{12}c^2_{13}c_{23}s_\delta 
\\
& \\
2 & {\bf R}_{23}(-\delta11){\bf R}_{12}(\delta11){\bf R}_{13} =
\left(
\begin{array}{ccc}
c_{12} c_{13} &
s_{12} e^{-i\delta} & 
c_{12} s_{13} \\
- s_{13} s_{23} - c_{13} c_{23} s_{12} e^{i\delta}  &
c_{12} c_{23} &
c_{13} s_{23} - s_{13} c_{23} s_{12} e^{i\delta} \\
-s_{13} c_{23} + c_{13} s_{23} s_{12} e^{i\delta} &
- c_{12} s_{23} &
c_{13} c_{23} + s_{13} s_{23} s_{12} e^{i\delta}
\end{array}
\right)
& 
s_{12}s_{13}s_{23}c^2_{12}c_{13}c_{23}s_\delta 
\\
& \\
3 & (-\delta11){\bf R}_{13}(\delta11){\bf R}_{23}{\bf R}_{12} =
\left(
\begin{array}{ccc}
c_{12} c_{23} + s_{12} s_{13} s_{23} e^{-i\delta}  &
s_{12} c_{13} - c_{12} s_{23} s_{13} e^{-i\delta} &
s_{13} c_{23} e^{-i\delta}\\
- s_{12} c_{23} &
c_{12} c_{23} & 
s_{23} \\
- c_{12} s_{13} e^{i\delta} + s_{12} s_{23} c_{13} & 
- s_{12} s_{13} e^{i\delta} - c_{12} s_{23} c_{13} &
c_{13} c_{23}
\end{array}
\right)
& 
s_{12}s_{13}s_{23}c_{12}c_{13}c^2_{23}s_\delta
\\
& \\
4 & (-\delta11){\bf R}_{12}(\delta11){\bf R}_{23}{\bf R}'_{12} =
\left(
\begin{array}{ccc}
c_{12} c'_{12} - s'_{12} s_{12} c_{23} e^{-i\delta} &
c_{12} s'_{12} + s_{12} c_{23} c'_{12} e^{-i\delta}   &
s_{12} s_{23} e^{-i\delta} \\
- s_{12} c'_{12} e^{i\delta} - s'_{12} c_{12} c_{23} &
- s_{12} s'_{12} e^{i\delta} + c_{12} c'_{12} c_{23}  &
c_{12} s_{23} \\ 
s'_{12} s_{23} &
- c'_{12} s_{23} &
c_{23}
\end{array}
\right)
& 
s_{12}s'_{12}s^2_{23}c_{12}c'_{12}c_{23}s_\delta
\\
& \\
5 & (-\delta11){\bf R}_{13}(\delta11){\bf R}_{23}{\bf R}'_{13} =
\left(
\begin{array}{ccc}
c_{13} c'_{13} - s'_{13} s_{13} c_{23} e^{-i\delta} &
- s_{13} s_{23} e^{-i\delta} &
c_{13} s'_{13} + s_{13} c_{23} c'_{13} e^{-i\delta}   \\
- s'_{13} s_{23} &
c_{23} &
c'_{13} s_{23} \\
- s_{13} c'_{13} e^{i\delta} - s'_{13} c_{13} s_{23} &
- c_{13} s_{23} &
- s_{13} s'_{13} e^{i\delta} + c_{13} c'_{13} c_{23}
\end{array}
\right)
& 
s_{13}s'_{13}s^2_{23}c_{13}c'_{13}c_{23}s_\delta
\\
& \\
6 & (11-\delta){\bf R}_{23}(11\delta){\bf R}_{12}{\bf R}'_{23} =
\left(
\begin{array}{ccc}
c_{12} &
s_{12} c'_{23} &
s_{12} s'_{23} \\
- s_{12} c_{23} &
c_{12} c_{23} c'_{23} - s_{23} s'_{23} e^{i\delta} &
c_{12} c_{23} s'_{23} + s_{23} c'_{23} e^{i\delta} \\
s_{12} s_{23} e^{-i\delta} &
- c_{12} c'_{23} s_{23} e^{-i\delta} - s'_{23} c_{23} &
- c_{12} s'_{23} s_{23} e^{-i\delta} + c_{23} c'_{23}
\end{array}
\right)
& 
s^2_{12} s_{23} s'_{23} c_{12} c_{23}c'_{23}s_\delta
\\
& \\
7 & (11-\delta){\bf R}_{13}(11\delta){\bf R}_{12}{\bf R}'_{13} =
\left(
\begin{array}{ccc}
c_{13} c'_{13} c_{12} - s_{13} s'_{13} e^{i\delta} &
c_{13} s_{12} &
c_{13} c_{12} s'_{13} + s_{13} c'_{13} e^{i\delta}  \\
- s_{12} c'_{13} &
c_{12} &
- s_{12} s'_{13} \\
- s_{13} c_{12} c'_{13} e^{-i\delta} - c_{13} s'_{13} &
- s_{13} s_{12} e^{-i\delta} &
- s_{13} s'_{13} c_{12} e^{-i\delta} + c_{13} c'_{13}
\end{array}
\right)
& 
s^2_{12}s_{13}s_{13}c_{12}c_{13}c'_{13}s_\delta
\\
& \\
8 & (11-\delta){\bf R}_{23}(11\delta){\bf R}_{13}{\bf R}'_{23} =
\left(
\begin{array}{ccc}
c_{13} &
- s_{13} s'_{23} &
s_{13} c'_{23} \\
- s_{13} s_{23} e^{-i\delta} &
c_{23} c'_{23} - s_{23} s'_{23} c_{13} e^{i\delta} &
c_{23} s'_{23} + s_{23} c_{13} c'_{23} e^{i\delta}  \\
- s_{13} c_{23} &
- s_{23} c'_{23} e^{-i\delta} - c_{23} c_{13} s'_{23} &
- s_{23} s'_{23} e^{-i\delta} + c_{13} c_{23} c'_{23}
\end{array}
\right)
& 
s^2_{13}s_{23}s'_{23}c_{13}c_{23}c'_{23}s_\delta
\\
& \\
9 & (1-\delta1){\bf R}_{12}(1\delta1){\bf R}_{13}{\bf R}'_{12} =
\left(
\begin{array}{ccc}
c_{12} c_{13} c'_{12} - s_{12} s'_{12} e^{i\delta} &
c_{12} c_{13} s'_{12} + s_{12} c'_{12} e^{i\delta}   &
c_{12} s_{13} \\
- s_{12} c_{13} c'_{12} e^{-i\delta} - s'_{12} c_{12} &
- s_{12} s'_{12} c_{13} e^{-i\delta} + c_{12} c'_{12}  &
- s_{12} s_{13} e^{-i\delta} \\ 
- s_{13} c'_{12} &
- s_{13} s'_{12} &
c_{13}
\end{array}
\right)
& 
s_{12}s'_{12}s^2_{13}c_{12}c'_{12}c_{13}s_\delta
\end{array}
\ee
\caption{Parametrizations of the Cabibbo-Kobayashi-Maskawa
matrix ${\bf K}$ in terms of three rotation
angles and one phase. ${\bf R}_{12}$ is the rotation in the 1-2 plane
by an angle $\theta_{12}$, $(\delta11)={\rm diag}(e^{i\delta},1,1)$, etc.
Three further parametrizations can be obtained by
taking the hermitian conjugate of the first three parametrizations.
$J$ is the Jarlskog invariant 
$J=|{\rm Im} (K_{ij} K_{kl} K_{il}^* K_{kj}^*)|$.
An arbitrary unitary matrix ${\bf V}$ is given by  
${\bf V} = {\rm diag} (1, e^{i\alpha_4}, e^{i\alpha_5} ) \,
{\bf K} \, {\rm diag} (e^{i \alpha_1}, e^{i\alpha_2}, e^{i\alpha_3})$
with $\alpha$s arbitrary. }
\end{table}

Three more parametrizations can be achieved by simply taking the hermitian
conjugate of the first three entries in Table I.
As in the example above, five phases will always
multiply from the far left and far right, and I call them $\alpha_i$
(i=1,..,5). 
Table I lists the middle matrix, ${\bf K}$, which is related to
${\bf V}$ by
\be
{\bf V} = 
(1 \alpha_4 \alpha_5 )
{\bf K} 
(\alpha_1 \alpha_2 \alpha_3) 
\ee
The Table lists also the Jarlskog invariant\cite{jarl85}
$J=|{\rm Im} (K_{ij} K_{kl} K_{il}^* K_{kj}^*)|$.

One angle is always between the rotations and I call it
$\delta$. Because of the free phases, $\delta$ can be moved
within the rotations. Here I adopt the following convention (although
any other parametrization is obtained by simple multiplication of phases
from left or right). $\delta$ and $-\delta$ are always around one rotation
through which they do not "go through", i.e. they cancel only if the 
rotation is trivial. In the example worked out and matrix 1 in Table I, 
the middle rotation is ${\bf R}_{13}$ so $\delta$ can be in the $1-1$ or
$3-3$ entry. Notice that we positioned the phase $\delta$ so that the
matrix containing ${-\delta}$ can always
be pulled out of ${\bf K}$ and reabsorbed into one of the $\alpha_i$.
Notice that the presence of both $\delta$ and $-\delta$ matrices 
in $\bf K$ is useful in seeing how the
complex phase $\delta$ explicitly dissapears when the rotation angle
between vanishes (for example in matrix 1 of the Table, when
$s_{13}=0$, $\delta$ also dissapears, and there is
no CP violation)\cite{wolf83,gron85}.
This convention however still leaves some freedom of where $\delta$
exactly appears in ${\bf K}$. The physical obsrervables, of
course do not depend on the choice of its position.

Each representation starts with choosing one entry to be just a 
single sine or cosine of an angle. In the example above $|V_{13}| \equiv
s_{13}$. I choose the convention if the
starting entry is off diagonal we choose a sine, and if it is diagonal we
choose a cosine. 

As I said before, the matrix K actually represents the most general
CKM since the phases $\alpha_i$ are arbitrary and can be set to
zero\footnote{The meaning of
the phases $\alpha_i$ is exactly the familiar freedom of redefinition of
the left handed up and down quark fields and then absorbed in the right
handed fields thus disappearing from the theory.}.
Some of the representations for the CKM already appeared in
the literature (up to a possible moving of the phase $\delta$ by phase
multiplications from left and right that are absorbed in $\alpha_i$s
and sign ambiguity in sinuses of the angles). 
Matrix 1 is the standard parametrization of Maiani and Chau and Keung
\cite{maia77,chau84}. Matrix 2 was proposed by Maiani in
\cite{maia76}. Matrix 4 is the parametrization obtained 
in \cite{dimo92}. 
Matrix 6 is the
Kobayashi-Maskawa parametrization \cite{koba73} (which in our notation
would be $(11,-1){\bf R}_{23}(11\delta){\bf R}^T_{12}{\bf R}'_{23}$).

Now we have everything ready for the construction of CKM directly from 
the quark mass matrices. I first start with the two generation case,
building up for the case of three generations which involves more
complicated matrix manipulations.

\vspace{0.2in}

{\bf D. Case of two  generations}\hspace{0.5cm} 

Here I do the diagonalization of a $2 \times 2$ matrix.
Let us get for example the form for ${\bf V}$ in terms of elements from
${\bf h}$. We see that 
\be 
{\bf h}^\dagger {\bf h} = 
\left(
\begin{array}{cc}
|h_{11}|^2 + |h_{21}|^2 & h_{11}^* h_{12}
+ h_{21}^* h_{22}\\
h_{11} h_{12}^*
+ h_{21} h_{22}^* & 
|h_{12}|^2 + |h_{22}|^2
\end{array}
\right)
\ee
which is of the form
\be 
{\bf h}^\dagger {\bf h} = 
\left(
\begin{array}{cc}
\lambda_{11} & \lambda_{12} e^{-i\gamma}\\
\lambda_{12} e^{i\gamma} & \lambda_{22}
\end{array}
\right)
\ee 
with real $\lambda_{ij}$ and $\gamma$ depending on the elements of 
${\bf h}$. Now we can first pull out the phase $\gamma$ and
then diagonalize
the real and symmetric matrix $\lambda_{ij}$ with one
rotation angle $\theta$ to get
\begin{eqnarray} 
{\bf h}^\dagger {\bf h} & = &
\left(
\begin{array}{cc}
1 & \\
& e^{i\gamma}
\end{array}
\right)
\left(
\begin{array}{cc}
c & s\\
-s & c
\end{array}
\right)
\left(
\begin{array}{cc}
e^{i\alpha} & \\
& e^{i\beta}
\end{array}
\right) 
\nonumber\\
& \times &
\left(
\begin{array}{cc}
m_1^2 & \\
 & m_2^2
\end{array}
\right)
\nonumber\\
& \times &
\left(
\begin{array}{cc}
e^{-i\alpha} & \\
& e^{-i\beta}
\end{array}
\right)
\left(
\begin{array}{cc}
c & -s\\
s & c
\end{array}
\right)
\left(
\begin{array}{cc}
1 & \\
& e^{-i\gamma}
\end{array}
\right)
\end{eqnarray}
where $\alpha$ and $\beta$ are the arbitrary phases. Also,
$m^2_{1,2} = 
{\lambda_{11}+\lambda_{22} \over 2} \mp
\sqrt{
({\lambda_{11}-\lambda_{22} \over 2})^2 + \lambda_{12}^2}$ 
and 
$ \tan \theta = s / c = { \lambda_{12} \over {m^2_2 - \lambda_{11}} }$.

Comparing this to 
(\ref{defineuv}) we see that ${\bf V}$ is equal to the first three
matrices on the right hand side and it is of the
general form for a unitary $2 \times 2$ unitary matrix as derived in
Section C
\be 
{\bf V} = 
\left(
\begin{array}{cc}
1 & \\
& e^{i\gamma}
\end{array}
\right)
\left(
\begin{array}{cc}
c & s\\
-s & c
\end{array}
\right)
\left(
\begin{array}{cc}
e^{i\alpha} & \\
& e^{i\beta}
\end{array}
\right) 
\ee
Similarly, one can construct ${\bf U}$. It is crucial to
remember that at this stage only the phases $\alpha$ and $\beta$ are
arbitrary, whereas $\gamma$ and the angle $\theta$ are fixed by the
original matrix ${\bf h}$. 

Now let us count the number of parameters. ${\bf h}$ has 4 real
and 4 imaginary parameters.
Compare this to the number of parameters in 
${\bf U}$ and ${\bf V}$.
Each one has 1 angle and 3 phases, which with the two mass eigenvalues
gives a total of 4 angles and 6 phases. Thus two combinations of
phases in ${\bf U}$ and ${\bf V}$ will not appear in ${\bf h}$. This
can be seen as follows. From
\be
{\bf h} = {\bf U} {\bf m} {\bf V}^\dagger
\ee
we see explicitly that of the 2 pairs or arbitrary phases
in ${\bf U}$ and ${\bf V}$ only two combinations appear, that is 
$\alpha_{\bf U}-\alpha_{\bf V}$
and
$\beta_{\bf U}-\beta_{\bf V}$. Two of the phases remain arbitrary
and in the next section I will choose them to be $\alpha_{\bf V}$ and
$\beta_{\bf V}$.

Now we have everything ready to construct the CKM matrix in the 2
generation cases, that we know consists of one angle only and no phases.

\vspace{0.2in}

{\bf E. CKM matrix for the two generations}\hspace{0.5cm} 

The $2 \times 2$ CKM matrix is given by
\ba
{\bf K} & = &{\bf V}^{u\dagger} {\bf V}^d = \nonumber\\
\times & &\left(
\begin{array}{cc}
e^{-i\alpha^u} & \\
& e^{-i\beta^u}
\end{array}
\right)
\left(
\begin{array}{cc}
c^u & -s^u \\
s^u & c^u 
\end{array}
\right)
\left(
\begin{array}{cc}
1 &  \\
 & e^{i\gamma} 
\end{array}
\right)
\left(
\begin{array}{cc}
c^d & s^d \\
-s^d & c^d 
\end{array}
\right)
\left(
\begin{array}{cc}
e^{i\alpha^d} & \\
& e^{i\beta^d}
\end{array}
\right) 
\ea
where $\gamma=\gamma^d-\gamma^u$. 

At this point $\alpha^i$ and $\beta^i$ are completely arbitrary
(independent of the original Yukawa matrices)
and I can use them to get rid of all the phases
For this purpose it is enough to notice that the three middle matrices on
the rhs can be written as

\be
\left(
\begin{array}{cc}
c^u & -s^u \\
s^u & c^u 
\end{array}
\right)
\left(
\begin{array}{cc}
1 &  \\
 & e^{i\gamma} 
\end{array}
\right)
\left(
\begin{array}{cc}
c^d & s^d \\
-s^d & c^d 
\end{array}
\right)
=
\left(
\begin{array}{cc}
c e^{i\phi} & 
s e^{i\xi} \\
- e^{i\gamma} s e^{-i\xi} & 
e^{i\gamma} c e^{-i\phi}
\end{array}
\right)=
\left(
\begin{array}{cc}
1 &  \\
  & e^{i(\gamma-\phi-\xi)} 
\end{array}
\right)
\left(
\begin{array}{cc}
c & s \\
-s & c 
\end{array}
\right)
\left(
\begin{array}{cc}
e^{i\phi} &  \\
& e^{i\xi} 
\end{array}
\right)
\label{combine}
\ee
where 
\be
s \equiv | c^u s^d - s^u c^d e^{i\gamma}|
\, ,\, 
c \equiv | c^u c^d + s^u s^d e^{i\gamma}|
\ee
are real and satisfy $c^2+s^2=1$, {\it i.e.} they
describe one angle. Also
$\xi \equiv \arg (c^u s^d - s^u c^d e^{i\gamma})$ and
$\phi \equiv \arg ( c^u c^d + s^u s^d e^{i\gamma})$.
We can now choose for example 
$\alpha^u = 0$,
$\beta^u = \gamma -\phi - \xi$,  
$\alpha^d = -\phi$,
$\beta^u = - \xi$  
so that indeed the CKM matrix in the 2 generation case is described by
one angle only and no phases
\be
{\bf K} =
\left(
\begin{array}{cc}
c & s  \\
- s & c
\end{array}
\right) \, .
\ee
It is interesting to note that although no CP violation is
present in the CKM for the two generations, 
the rotation angle depends on the complex phases in
the quark mass matrices.

{\bf Example.}
Let us assume that the mass matrix in the down sector
is of the form \cite{frit77}
\be
{\bf h} =
\left(
\begin{array}{cc}
0 & B e^{-i\gamma}  \\
B e^{i\gamma} & A
\end{array}
\right) \, ,
\ee
and that there is a similar structure in the up sector.
The mass eigenvalues and the rotation angle are 
\be
m_{d,s} = | A/2 \mp \sqrt{A^2/4 + B^2}| \, , \,
\tan\theta^d = { B \over m_s} = \sqrt {m_d \over m_s} 
\ee
and similarly for the up sector
\footnote{The absolute values in the masses can be obtained
by actually diagonalizing ${\bf h}^\dagger {\bf h}$.}.
The CKM angle is given by \cite{frit79} 
\be
\sin \theta = |c^u s^d - s^u c^d e^{i\gamma}| =
|\sqrt {m_d \over m_s} - \sqrt {m_u \over m_c} e^{i\gamma}|
{ 1 \over \sqrt {1 + {m_d \over m_s} } } 
{ 1 \over \sqrt {1 + {m_u \over m_c} } } 
\ee
where $\gamma = \gamma^d - \gamma^u$. These results are {\it exact}.

\vspace{0.2in}

{\bf F. Case of three generations}\hspace{0.5cm} 

Let us now do the more nontrivial case of three generations.
Let us again get the form for ${\bf V}$ in terms of elements of ${\bf h}$.
${\bf h}^\dagger {\bf h}$ is of the form
\be 
{\bf h}^\dagger {\bf h} = 
\left(
\begin{array}{ccc}
\lambda_{11} & \lambda_{12} e^{-i\gamma_{12}} & \lambda_{13}
e^{-i\gamma_{13}}\\
\lambda_{12} e^{i\gamma_{12}} & \lambda_{22} & 
\lambda_{23} e^{-i\gamma_{23}}
\\
\lambda_{13} e^{i\gamma_{13}} & 
\lambda_{23} e^{i\gamma_{23}} & \lambda_{33}
\end{array}
\right)
\ee 
with real $\lambda_{ij}$ and $\gamma_{ij}$ depending on the elements of 
${\bf h}$. 
I will do the diagonalization by applying successive rotations and phase
redefinitions. Various parametrizations of the unitary matrices that
diagonalize ${\bf h}$ will depend on
which order of phase redefinitions and rotations in the process of
diagonalization we choose. 

I will now continue with a specific choice but
the steps can easily be repeated for any other
representation. The choice is inspired by the often assumed 
hierarchical structure in the quark mass matrices with the biggest
element at the 3-3 entry (and the other elements getting smaller
and smaller as we move away from the 3-3 entry), so that one can easily
follow
the results below by assuming that the 1-3 rotations are small.  
However, needless to stress, the results below are exact and
do not depend on any assumption.

In order to pull out all the phases I will need to do at least one
rotation between  phase redefinitions. First let me 
pull out the phase $\gamma_{13}$ and do the rotation ${\bf R}'_{13}$ in
the 1-3 sector to set the corresponding element to zero, and then pull out
the remaining phases
\begin{eqnarray}
{\bf h}^\dagger {\bf h} & = &
\left(
\begin{array}{ccc}
1 &  &  \\
& 1 & \\
& & e^{i\gamma_{13}} 
\end{array}
\right)
\left(
\begin{array}{ccc}
c'_{13} &  & s'_{13} \\
& 1 & \\
-s'_{13} &  & c_{13} 
\end{array}
\right)
\left(
\begin{array}{ccc}
e^{i\gamma'_{12}} &  &  \\
& 1 & \\
& & e^{i\gamma'_{23}} 
 \end{array}
\right) \nonumber\\
& \times &
\left(
\begin{array}{ccc}
\lambda'_{11} & \lambda'_{12} & 0 \\
\lambda'_{12} & \lambda_{22} & \lambda'_{23} \\
0 & \lambda'_{23} & \lambda'_{33}
\end{array}
\right) \nonumber\\
& \times &
\left(
\begin{array}{ccc}
e^{-i\gamma'_{12}} &  &  \\
& 1 & \\
& & e^{-i\gamma'_{23}} 
\end{array}
\right)
\left(
\begin{array}{ccc}
c'_{13} &  & - s'_{13} \\
& 1 & \\
s'_{13} &  & c_{13} 
\end{array}
\right)
\left(
\begin{array}{ccc}
1 &  &  \\
& 1 & \\
& & e^{-i\gamma_{13}} 
\end{array}
\right)
\end{eqnarray} 
where $\lambda'_{11},\lambda'_{33} = {{\lambda_{11}+\lambda_{33}} \over 2}
\mp \sqrt { ({{\lambda_{11}-\lambda_{33}} \over 2 })^2 
+ \lambda^2_{13}}$,
$\tan\theta'_{13} 
= { \lambda_{13} \over {\lambda'_{33} - \lambda_{11}} }$,
$\lambda'_{23} e^{i\gamma'_{23}} = 
\lambda_{12} s'_{13} e^{-i\gamma_{12}} + 
\lambda_{23} c'_{13} e^{i\gamma_{23}-i\gamma_{13}}$
and 
$\lambda'_{12} e^{-i\gamma'_{12}} = 
\lambda_{12} c'_{13} e^{i\gamma_{12}} - 
\lambda_{23} s'_{13} e^{-i\gamma_{23}+i\gamma_{13}}$.

We are now left with  diagonalizing the middle matrix, which I do in 
the Appendix A. Contrary to a recent claim\cite{frit97}, a general real
symmetric matrix with two off diagonal zeroes {\it cannot} be diagonalized
with two rotations only. If one chooses, for example, the 2-3 rotation to
get rid off  $\lambda'_{23}$ there will be a term 
$\lambda'_{12} s_{23}$ generated in the 1-3 position.
Alternatively, one may think that
by cleverly choosing two rotations one can simultaneously completely
diagonalize the matrix. This however can easily be seen to fail unless we
already start with the trivial case of, say, $\lambda'_{12}=0$. 

Thus, we need three rotations to diagonalize the middle matrix.
There is no rule in which order the three rotations should be taken.
I pick the first rotation to be a 1-3 rotation ${\bf R}''_{13}$ just 
so it is easy to combine later with ${\bf R}'_{13}$. My choice of the
subsequent rotations being  ${\bf R}_{23}$ and ${\bf R}_{12}$ is inspired
again by the hierarchical structure. So ${\bf h}^\dagger {\bf h}$ becomes
\begin{eqnarray}
{\bf h}^\dagger {\bf h} & = &
\left(
\begin{array}{ccc}
1 &  &  \\
& 1 & \\
& & e^{i\gamma_{13}} 
\end{array}
\right)
{\bf R}'_{13}
\left(
\begin{array}{ccc}
e^{i\gamma'_{12}} &  &  \\
& 1 & \\
& & e^{i\gamma'_{23}} 
 \end{array}
\right) 
{\bf R}''_{13} {\bf R}_{23} {\bf R}_{12}
\left(
\begin{array}{ccc}
e^{i\alpha_1} &  &  \\
& e^{i\alpha_2} & \\
& & e^{i\alpha_3} 
\end{array}
\right)
\nonumber\\
& \times &
\left(
\begin{array}{ccc}
m^2_1 & 0 & 0 \\
0 & m^2_2 & 0 \\
0 & 0 & m^2_3
\end{array}
\right) \nonumber\\
& \times &
\left(
\begin{array}{ccc}
e^{-i\alpha_1} &  &  \\
& e^{-i\alpha_2} & \\
& & e^{-i\alpha_3} 
\end{array}
\right)
{\bf R}^T_{12} {\bf R}^T_{23} {\bf R}''^T_{13}
\left(
\begin{array}{ccc}
e^{-i\gamma'_{12}} &  &  \\
& 1 & \\
& & e^{-i\gamma'_{23}} 
 \end{array}
\right)
{\bf R}'^T_{13}
\left(
\begin{array}{ccc}
1 &  &  \\
& 1 & \\
& & e^{-i\gamma_{13}} 
\end{array}
\right)
\end{eqnarray} 
in the obvious notation for the rotations. It is easy to obtain the
relations between the rotation angles and the matrix elements and we
leave the details to Appendix A. The phases $\alpha_i$ are the
arbitrary phases. We can now identify the unitary matrix ${\bf V}$
as the first line on the rhs of the above equation
\be
{\bf V} =
\left(
\begin{array}{ccc}
e^{i\gamma'_{12}}  &  &  \\
& 1 & \\
& & e^{i\gamma_{13}+i\gamma'_{12}} 
\end{array}
\right)
{\bf R}'_{13}
\left(
\begin{array}{ccc}
1  &  &  \\
& 1 & \\
& & e^{i\gamma} 
\end{array}
\right) 
{\bf R}''_{13} {\bf R}_{23} {\bf R}_{12}
\left(
\begin{array}{ccc}
e^{i\alpha_1} &  &  \\
& e^{i\alpha_2} & \\
& & e^{i\alpha_3} 
\end{array}
\right)
\ee
where $\gamma=\gamma'_{23}-\gamma'_{12}$. 
We can bring the matrix into one of the forms
for unitary matrices given in Table I. First, I
combine the two 1-3 rotations with the phase
$\gamma$ between into one 1-3
rotation, between two phases, similarly to equation (\ref{combine}):
\be
{\bf R}'_{13} (11\gamma) {\bf R}''_{13} =
(11,\gamma-\phi-\xi){\bf R}_{13} (\phi1\xi) =
(\xi1,\gamma-\phi){\bf R}_{13} (\phi-\xi,11) 
\label{onethreerots}
\ee
where $s_{13} = 
|c'_{13}s''_{13}+s'_{13}c''_{13}e^{i\gamma}|$,
$\xi = \arg (c'_{13}s''_{13}+s'_{13}c''_{13}e^{i\gamma})$
and
$\phi = \arg (c'_{13}c''_{13}-s'_{13}s''_{13}e^{i\gamma})$.
After a further trivial phase redefinition we arrive at the form of
a general arbitrary unitary matrix listed in Table I (matrix 3),
\ba
{\bf V} & = &
(1\alpha_4\alpha_5)(-\delta11) {\bf R}_{13}(\delta11){\bf
R}_{23}{\bf R}_{12} (\alpha_1\alpha_2\alpha_3) = \nonumber\\
& & 
\left(
\begin{array}{ccc}
1 &  &  \\
& e^{i\alpha_4} & \\
& & e^{i\alpha_5} 
\end{array}
\right)
\left(
\begin{array}{ccc}
c_{12} c_{23} + s_{12} s_{13} s_{23} e^{-i\delta}  &
s_{12} c_{13} - c_{12} s_{23} s_{13} e^{-i\delta} &
s_{13} c_{23} e^{-i\delta}\\
- s_{12} c_{23} &
c_{12} c_{23} & 
s_{23} \\
- c_{12} s_{13} e^{i\delta} + s_{12} s_{23} c_{13} & 
- s_{12} s_{13} e^{i\delta} - c_{12} s_{23} c_{13} &
c_{13} c_{23}
\end{array}
\right)
\left(
\begin{array}{ccc}
e^{i\alpha_1} &  &  \\
& e^{i\alpha_2} & \\
& & e^{i\alpha_3} 
\end{array}
\right)
\ea
where $\delta = \phi - \xi$, $\alpha_4 = -\phi - \gamma'_{12}$,
$\alpha_5 = \gamma_{13} + \gamma - 2 \phi$.

As expected the unitary matrix ${\bf V}$ depends on 3 angles and
6 phases. Of the six phases only three of $\alpha_i$ are 
arbitrary ($\alpha_1,\alpha_2,\alpha_3$),
and others are fixed by the matrix $\bf h$.
Notice of course the ambiguity in placing the $\alpha$s on the left and
right end, as well as the ambiguity in placing $\delta$ between the
matrices.

The counting of parameters proceeds similarly to the $2 \times 2$ case.
From ${\bf h} = {\bf U} {\bf m} {\bf V}^\dagger$ we see that
the 9 real parameters on the lhs are matched on the rhs by the 
3 angles in ${\bf U}$, 3 eigenvalues in ${\bf m}$ and
3 angles in ${\bf V}$. For the phases, ${\bf h}$ has 9 of them.
On the other side, ${\bf U}$ and ${\bf V}$ have six phases each.
However, three combinations of the six phases that were arbitrary (3 in
${\bf U}$ and 3 in ${\bf V}$) do not enter the rhs. I will again choose
these arbitrary phases to be the $\alpha_1,\alpha_2,\alpha_3$ in ${\bf V}$
when we consider the CKM in the next Section.

\vspace{0.2in}

{\bf G. CKM matrix for the three generations}\hspace{0.5cm}

I choose the parametrization derived in the previous section
(but any other is equally good) for the two left handed rotations
\be
{\bf V}^i = (1\alpha^i_4\alpha^i_5)
(-\delta^i11){\bf R}^i_{13} (\delta^i11) {\bf R}^i_{23} {\bf
R}^i_{12}
(\alpha^i_1\alpha^i_2\alpha^i_3)
\ee
where $i=u,d$. Only the phases $\alpha^i_1,\alpha^i_2,\alpha^i_3$ 
do not depend on the original quark mass matrices. CKM is given by
\be
{\bf K} = {\bf V}^{u\dagger} {\bf V}^d =
(-\alpha^u_1-\alpha^u_2-\alpha^u_3){\bf R}^{uT}_{12}{\bf
R}^{uT}_{23} 
(-\delta_d,\alpha_4^d-\alpha_4^u,\delta^u-\delta^d)
{\bf R}^{uT}_{13}
(11\gamma)
{\bf R}^d_{13}
(\delta^d11){\bf R}^d_{23}{\bf R}^d_{12}
(\alpha^d_1\alpha^d_2\alpha^d_3)
\label{ckmphase}
\ee
where $\gamma=\delta^d - \delta^u + \alpha_5^d- \alpha_5^u  $.

This expression can be transformed into one of the parametrizations of
the CKM with three rotations and one phase. 
We can arrive at any of the nine parametrizations given in Table I, 
but I choose the one requiring the least number of manipulations, the
parametrization 4, and I derive it here. For this purpose I will use
manipulations with rotations and as well the freedom in $\alpha$s.
However, since the manipulations of rotations involve phases between
them, and keeping track becomes more cumbersome,
I first derive the CKM with no phases.
Then I consider the most general case of nonvanishing phases.

{\bf $3\times3$ CKM with no phases}

In this case the CKM from (\ref{ckmphase}) is given by
\be
{\bf K} = {\bf V}^{u\dagger} {\bf V}^d =
{\bf R}^{uT}_{12}
{\bf R}^{uT}_{23} {\bf R}^{uT}_{13}{\bf R}^d_{13}
{\bf R}^d_{23}{\bf R}^d_{12}
\ee
First, notice that the product ${\bf R}^{uT}_{13}{\bf R}^d_{13}$ can
be written as one 1-3 rotation ${\bf R_{13}}$ with angle
$\theta_{13}=\theta^d_{13}-\theta^u_{13}$.
Then, we can find rotations
${\bf R}''_{12}$, ${\bf R}_{23}$, ${\bf R}'''_{12}$, such that
\be
{\bf R}^{uT}_{23} {\bf R}_{13}{\bf R}^d_{23}
=
{\bf R}''_{12} {\bf R}_{23}{\bf R}'''_{12}
\ee 
where we read the relations between the angles from the (very useful!)
Table I, comparing matrices 4 and 8, with all phases set to zero
\be
c_{23}= c_{13} c^u_{23} c^d_{23} + s^u_{23} s^d_{23} \, , \,
s''_{12} s_{23} = s_{13} c^d_{23} \, , \,
s'''_{12} s_{23} = - s_{13} c^u_{23} 
\ee
Finally, combining the two pairs of 1-2 rotations 
(${\bf R}^{uT}_{12}$ and ${\bf R}''_{12}$, 
${\bf R}^d_{12}$ and ${\bf R}'''_{12}$) into just two 1-2 rotations
(${\bf R}_{12}$ and ${\bf R}'_{12}$ respectively)
with 
$\theta_{12} = \theta''_{12} - \theta^u_{12}$ 
and 
$\theta'_{12} = \theta'''_{12} + \theta^d_{12}$ 
we get
\be
{\bf K} = {\bf R}_{12} {\bf R}_{23} {\bf R}'_{12}
=
\left(
\begin{array}{ccc}
c_{12} c'_{12} - s'_{12} s_{12} c_{23}  &
c_{12} s'_{12} + s_{12} c_{23} c'_{12}  &
s_{12} s_{23}  \\
- s_{12} c'_{12} - s'_{12} c_{12} c_{23} &
- s_{12} s'_{12} + c_{12} c'_{12} c_{23}  &
c_{12} s_{23} \\ 
s'_{12} s_{23} &
- c'_{12} s_{23} &
c_{23}
\end{array}
\right)
\ee

{\bf $3\times3$ CKM with phases}

Now let me consider the most general case in equation
(\ref{ckmphase}). The product of the two 1-3 rotations with
the phase $\gamma$ between can be
combined into one 1-3 rotation ${\bf R}_{13}$ between two phase
transformations,  similar to equation (\ref{combine})
or (\ref{onethreerots}). At this point we have
\be
{\bf K} =
(-\alpha^u_1-\alpha^u_2-\alpha^u_3){\bf R}^{uT}_{12}
(-\delta^d+\xi,\alpha^d_4-\alpha^u_4,\alpha^d_4-\alpha^u_4)
{\bf R}^{uT}_{23}
(11\rho)
{\bf R}_{13} {\bf R}^d_{23}
(\delta^d+\phi-\xi,1,1)
{\bf R}^d_{12}
(\alpha^d_1\alpha^d_2\alpha^d_3)
\ee
where $\rho = \alpha^d_5 - \alpha^u_5 - (\alpha^d_4 - \alpha^u_4) - \phi$.
Now we can recognize the product of the middle three rotations
as one of the entries in Table I (matrix 8). We can
immediately write it then as matrix 4, with appropriate phases
on the left and right, since we relate two forms of the same unitary
matrix. We call the phases $\beta_i$, $i=1,...,5$ rather then 
$\alpha_i$, to point out that they depend on the initial quark mass
matrices
\be
{\bf R}^{uT}_{23} (11\rho) {\bf R}_{13} {\bf R}^d_{23}
=
(1\beta_4\beta_5)
{\bf R}''_{12} (\delta'11) {\bf R}_{23} {\bf R}'''_{12}
(\beta_1\beta_2\beta_3)
\ee
Comparing in Table I we get
\ba
c_{23} & = & |c_{13} c^u_{23} c^d_{23} e^{i\rho} + s^u_{23} s^d_{23}| \, ,
\, s''_{12} s_{23} = s_{13} c^d_{23} \, , \,
s'''_{12} s_{23} = - s_{13} c^u_{23} \nonumber\\
\beta_3 & = & 0 \, , \, 
\beta_5 = \arg(s^u_{23} s^d_{23} + c_{13} c^u_{23}
c^d_{23} e^{\i\rho}) 
\, , \, \beta_1 = \rho - \beta_5 \nonumber\\
\beta_4 & = & \arg(c^u_{23} s^d_{23} - c_{13} c^d_{23}
s^u_{23} e^{\i\rho}) \, , \,
\beta_2 = -\beta_5 + \arg(s^u_{23} c^d_{23} - c_{13} s^d_{23}
c^u_{23} e^{\i\rho}) \, , \,
\delta' = \arg(c_{13} e^{i\beta_1} +  s''_{12} s'''_{12}) c_{23})
\, , \,
\ea 

So we write
\be
{\bf K} =
(-\alpha^u_1-\alpha^u_2-\alpha^u_3){\bf R}^{uT}_{12}
(1\delta''1)
{\bf R}''_{12}(\delta'11){\bf R}_{23} {\bf R}'''_{12}
(1\delta'''1) {\bf R}^d_{12}
(\alpha^d_1\alpha^d_2\alpha^d_3)
\ee
where most of the phases were trivially absorbed into $\alpha_i$
and
$\delta''=\delta^d+\alpha^d_4-\alpha^u_4+\beta_4-\xi$
and
$\delta'''=-\delta^d+\beta_2-\beta_1-\phi+\xi$. 
What is left is to combine the two pairs of 1-2 rotations 
(${\bf R}^{uT}_{12}$ and ${\bf R}''_{12}$, 
${\bf R}^d_{12}$ and ${\bf R}'''_{12}$) with
the phases between just two 1-2 rotations (${\bf R}_{12}$ and
${\bf R}'_{12}$) between phases, similar to equation (\ref{combine}).

And finally I can choose $\alpha_i$ to rotate away
all phases that depend on the original quark mass matrices, except
one which we call $\delta$. So I arrive at the form for the CKM
\be
{\bf K} =
(-\delta11){\bf R}_{12}(\delta11){\bf R}_{23}{\bf R}'_{12} =
\left(
\begin{array}{ccc}
c_{12} c'_{12} - s'_{12} s_{12} c_{23} e^{-i\delta} &
c_{12} s'_{12} + s_{12} c_{23} c'_{12} e^{-i\delta}   &
s_{12} s_{23} e^{-i\delta} \\
- s_{12} c'_{12} e^{i\delta} - s'_{12} c_{12} c_{23} &
- s_{12} s'_{12} e^{i\delta} + c_{12} c'_{12} c_{23}  &
c_{12} s_{23} \\ 
s'_{12} s_{23} &
- c'_{12} s_{23} &
c_{23}
\end{array}
\right)
\ee
which corresponds to matrix 4 in Table I.

As stressed before, there are essentially twelve different 
parametrizations of the CKM that are given in Table I,
and we can arrive at any of them. It is just a matter of picking the
order of rotations in the process of diagonalization, and manipulations in
the CKM to get to any of the standard forms with three rotation angles
and one phase.

Although maybe cumbersome, the above results are exact. 
It is instructive to repeat the above excercise approximately for
the often assumed case of hierarchical quark mass matrices, where
the largest element is $h_{33}$, and the elements get smaller as
we get farther away from this element. Then the angles in the CKM
parametrization derived above (matrix 4 in Table 1) are 
in simple approximate relation with the elements of the Yukawa matrices.
For this purpose let us first look at the derivation of the
diagonalization matrices ${\bf V}^i$, i=u,d,  in Section C.
 First, as advertised before, let me show that we can in first
approximation neglect the 1-3 rotations.
We see that because
of the assumed hierarchy 
$\theta^{i'}_{13} \approx \lambda^i_{13} / \lambda^i_3 
\approx (h^i_{21}h^i_{23}+h^i_{31}h^i_{33})/h^{i2}_{33}$ is very small.
Also from Appendix A, 
$\theta^{i''}_{13} \approx \lambda^{i'}_{23} \lambda^{i'}_{12}/
\lambda^{i2}_3$
is very small. Now, from Appendix A,
$\theta_{23} \approx \lambda^i_{23} / \lambda^i_3 \approx h^i_{32} /
h^i_{33}$
and 
$\theta^i_{12} \approx \lambda^i_{12} / \lambda^i_2 
\approx (h^i_{21} h^i_{22} + h^i_{31} h^i_{32}) / m^{i2}_2$.
Now, let us look at CKM. Since we neglect 1-3 rotations,
${\bf R}''_{12}$ and ${\bf R}'''_{12}$ can be neglected, and
$\theta_{23} \approx | \theta^d_{23} - \theta^u_{23} e^{i\rho}|$, while
$\theta_{12} \approx -\theta^u_{12}$
and
$\theta'_{12} \approx \theta^d_{12}$.

{\bf Example.}
Let us assume that the up and down quark mass matrices are of the
form\cite{geor79,heho90,dimo92}
\be
{\bf h}^u =
\left(
\begin{array}{ccc}
0 & C & 0 \\
C & 0 & B \\
0 & B & A 
\end{array}
\right) \, , \,
{\bf h}^d =
\left(
\begin{array}{ccc}
0 & F & 0 \\
F & E & 0 \\
0 & 0 & D 
\end{array}
\right)
\ee
with all the nonzero entries complex in principle.
I can trivially pull out their phases
\be
{\bf h}^u =
\left(
\begin{array}{ccc}
e^{i\gamma_C-i\gamma_B} & &\\
& e^{i\gamma_B-i\gamma_A}& \\
& & 1 
\end{array}
\right)
\left(
\begin{array}{ccc}
0 & C & 0 \\
C & 0 & B \\
0 & B & A 
\end{array}
\right)
\left(
\begin{array}{ccc}
e^{i\gamma_C-i\gamma_B+i\gamma_A} & &\\
& e^{i\gamma_B}& \\
& & e^{i\gamma_A} 
\end{array}
\right)
\ee
and similarly for ${\bf h}^d$
\be
{\bf h}^d =
\left(
\begin{array}{ccc}
e^{i\gamma_F-i\gamma_E} & &\\
& 1& \\
& & 1 
\end{array}
\right)
\left(
\begin{array}{ccc}
0 & F & 0 \\
F & E & 0 \\
0 & 0 & D 
\end{array}
\right)
\left(
\begin{array}{ccc}
e^{i\gamma_F} & &\\
& e^{i\gamma_E}& \\
& & e^{i\gamma_D} 
\end{array}
\right)
\ee
where $A,B,C,D,E,F$ are now real and positive. 

${\bf h}^d$ is now trivially diagonalized with one rotation
${\bf R}^d_{12}$
\begin{eqnarray}
{\bf h}^d & = &
\left(
\begin{array}{ccc}
e^{i\gamma_F-i\gamma_E} & &\\
& 1& \\
& & 1 
\end{array}
\right)
{\bf R}^d_{12}
\left(
\begin{array}{ccc}
e^{i\alpha_1} &  &  \\
& e^{i\alpha_2} & \\
& & e^{i\alpha_3} 
\end{array}
\right)
\nonumber\\
& \times &
\left(
\begin{array}{ccc}
m_d & 0 & 0 \\
0 & m_s & 0 \\
0 & 0 & m_b
\end{array}
\right) \nonumber\\
& \times &
\left(
\begin{array}{ccc}
e^{-i\beta_1} &  &  \\
& e^{-i\beta_2} & \\
& & e^{-i\beta_3} 
\end{array}
\right)
{\bf R}^{dT}_{12}
\left(
\begin{array}{ccc}
e^{i\gamma_F} & &\\
& e^{i\gamma_E} & \\
& & e^{i\gamma_D} 
\end{array}
\right)
\end{eqnarray} 
where the rotation angle is given by 
\be
\tan \theta^d_{12} = {F \over m_s} = \sqrt{ m_d \over m_s}.
\label{eq1}
\ee
In order to get the CKM as easy as possible I choose to diagonalize
the up quark matrix with the 1-2 2-3 1-2 rotations
\be
\left(
\begin{array}{ccc}
0 & C & 0 \\
C & 0 & B \\
0 & B & A 
\end{array}
\right)
=
{\bf R}^{u'}_{12} 
{\bf R}^u_{23} 
{\bf R}^u_{12} 
\left(
\begin{array}{ccc}
e^{i\alpha_1} &  &  \\
 & e^{i\alpha_2} &  \\
 &  & e^{i\alpha_3} 
\end{array}
\right)
\left(
\begin{array}{ccc}
m^2_u &  &  \\
 & m^2_c &  \\
 &  & m^2_t 
\end{array}
\right)
\left(
\begin{array}{ccc}
e^{-i\beta_1} &  &  \\
 & e^{-i\beta_2} &  \\
 &  & e^{-i\beta_3} 
\end{array}
\right)
{\bf R}^{uT}_{12} 
{\bf R}^{uT}_{23} 
{\bf R}^{u'T}_{12} 
\ee
where, using the matrix 4 in Table I with all phases set to
zero, 
\be
t^{u'}_{12} = {C \over m_t} \, , \, 
t^u_{23} ={ {AB} \over {m^2_t - B^2 - C^2} } { 1 \over c^{u'}_{12}}
\, , \, 
t^u_{12} = - { {BC} \over {m^2_c - C^2} } { s^u_{23} \over c^{u'}_{12} }
- t^{u'}_{12} c^u_{23}
\label{rots1}
\ee
and I have used the shorthand $t \equiv s/c$. The mass squares
$m^2_{u,c,t}$ are the solutions of the cubic equation
$\det ({\bf h}^\dagger {\bf h} - m^2 {\bf 1}) =
(C^2 - m^2) [ (B^2 + C^2 - m^2) (A^2 + B^2 - m^2) - A^2 B^2 ]
- B^2 C^2 ( B^2 + C^2 - m^2) = 0$. This can be written
as $(B^2 + C^2 - m^2)^2 m^2 = (C^2 - m^2)^2 A^2$, reflecting
the fact that one can also look for eigenvalues of ${\bf h}^u$
itself, but with all eigenvalues positive. Also, notice
$m_u m_c m_t = C^2 A$.

Now, we can easily read the matrices ${\bf V}^{u,d}$ and compute the 
CKM
\be
{\bf K} = {\bf V}^{u\dagger} {\bf V}^d =
\left(
\begin{array}{ccc}
e^{-i\alpha^u_1} &  &  \\
 & e^{-i\alpha^u_2} &  \\
 &  & e^{-i\alpha^u_3} 
\end{array}
\right)
{\bf R}^{uT}_{12} 
{\bf R}^{uT}_{23} 
{\bf R}^{u'T}_{12} 
\left(
\begin{array}{ccc}
1 & &\\
& e^{i\gamma}& \\
& & 1 
\end{array}
\right)
{\bf R}^d_{12}
\left(
\begin{array}{ccc}
e^{i\beta} & &\\
& e^{i\beta} & \\
& & 1 
\end{array}
\right)
\left(
\begin{array}{ccc}
e^{i\alpha^d_1} &  &  \\
& e^{i\alpha^d_2} & \\
& & e^{i\alpha^d_3} 
\end{array}
\right)
\ee
where $\beta =\gamma_C-\gamma_B+\gamma_F-\gamma_E$ and
$\gamma =  2\gamma_B-\gamma_A-\gamma_C-\gamma_F+\gamma_E$, and $\alpha$s
are the arbitrary phases. Again, we can combine the two 1-2
rotations with the phase $\gamma$ into one 1-2 rotation
just as in equation (\ref{combine}),
${\bf R}^{u'T}_{12} (1\gamma1) {\bf R}^d_{12} =
(1,\gamma-\phi-\xi,1) {\bf R}'_{12} (\phi\xi1)$,
where 
\be
s'_{12} = |c^{u'}_{12} s^d_{12} + c^d_{12} s{u'}_{12} e^{i\gamma}|
\label{rots2}
\ee
The freedom in $\alpha$s can be used to rotate away all phases except one.
this brings the CKM to the form of matrix 4 in Table I
\be
{\bf K} = (-\delta11) {\bf R}_{12} (\delta11)
{\bf R}_{23} {\bf R}'_{12}
\ee
where $\delta = -\gamma +\phi +\xi$, 
$\theta_{12} = -\theta^u_{12}$ and
$\theta_{23} = -\theta^u_{23}$.
 The rotation angles are given
in (\ref{eq1}), (\ref{rots1}) and (\ref{rots2}). This completes the {\it
exact} solution of the example.

One can check the approximate results previously obtained.
The cubic equation is solved approximately for $m_t \approx A$,
$m_c \approx B^2/ A$ and $m_u \approx C^2 / m_c$.
The mixing angles are $t^{u'}_{12} =  \sqrt{m_u m_c} / m_t$ (exact!),
$t^u_{23} \approx \sqrt{m_c / m_t}$, $t_{12} \approx C / m_c = 
\sqrt{ m_u \over m_c}$, and $t^d_{12} = \sqrt{m_d \over m_s}$ (exact!).
We see that $t''_{12}$ is very small compared to $t^d_{12}$ and for all
practical purposes one can take $t'_{12} \approx t^d_{12}$.
So we have to leading order for the CKM elements
$K_{cb} \approx K_{ts} \approx s_{23} \approx {B \over m_t} \approx
\sqrt{ {m_c} \over {m_t} }$,
$|{K_{ub} \over K_{cb}}| = |t_{12}| \approx \sqrt {m_u \over m_c}$ and
$|{K_{td} \over K_{ts}}| = |t'_{12}| \approx \sqrt {m_d \over m_s}$.

\vspace{0.2in}

{\bf H. Conclusions}

In conclusion, CKM can be parametrized in terms of three rotation angles
and one phase, but it is not clear which order of rotations and
positioning of the phase is best to use in a given
model of quark masses.  The problem is that one must construct 
the CKM from the unitary rotations on
the left handed quark fields. Each of the left up and down diagonalization
matrices consists of three angles and six phases and one must combine them
in a nontrivial way to obtain the CKM in a parametrization with three
angles and one phase. There are essentially 12 possible parametrizations
(up to overall phase multiplications) and I list them in Table I.

Some forms may turn out to be more practical than others. 
For example, if the elements of the quark mass matrices exhibit a
hierarchy, then the CKM parametrization as matrix 4 of Table I
seems to be the one that is most easily obtained from the diagonalization
matrices. In particular, this form has simple elements in the
top or bottom entries and may prove convenient in the analysis of heavy
quark processes \cite{dimo92,barb96,frit97}. 

\vspace{0.2in}

{\bf Acknowledgements}

The counting in Section B was inspired by Howard Haber's lectures
at the '97 ICTP Summer School and I thank him for a discussion.  
I thank Lawrence Hall and Atsushi Yamada for many useful comments and 
interesting discussions.

\vspace{0.2in}

{\bf Appendix A. Diagonalization of a real symmetric matrix with
two texture zeroes}

We want to diagonalize the matrix 
\be
{\bf M^2} = 
\left(
\begin{array}{ccc}
\lambda_{11} & \lambda_{12} & 0 \\
\lambda_{12} & \lambda_{22} & \lambda_{23} \\
0 & \lambda_{23} & \lambda_{33}
\end{array}
\right) 
\ee
The angles of the necessary rotations are then determined in terms
of the $\lambda_{ij}$ and the eigenvalues $\lambda_i$, $i=1,2,3$. 
The eigenvalues $\lambda_i$, are the
solutions of the cubic equation ${\rm det} ({\bf M}^2 -
\lambda {\bf 1})$
\be
(\lambda_{11}-\lambda)
[(\lambda_{22}-\lambda)(\lambda_{33}-\lambda)-\lambda^2_{23}]
-
(\lambda_{33}-\lambda)\lambda^2_{12} = 0
\ee
and we order them so that $\lambda_1$ is smallest and $\lambda_3$
largest.

As stressed in the
text, we need three rotations in order to diagonalize this matrix.
We choose the following order (although any other is equally good)
\be
{\bf M^2} = 
\left(
\begin{array}{ccc}
\lambda_1 & 0 & 0 \\
0 & \lambda_2 & 0 \\
0 & 0 & \lambda_3
\end{array}
\right)
=
{\bf R}^T_{12} {\bf R}^T_{23} {\bf R}^T_{13}
\left(
\begin{array}{ccc}
\lambda_{11} & \lambda_{12} & 0 \\
\lambda_{12} & \lambda_{22} & \lambda_{23} \\
0 & \lambda_{23} & \lambda_{33}
\end{array}
\right) 
{\bf R}_{13} {\bf R}_{23} {\bf R}_{12} 
\ee
We can rewrite the above equation
\be
\left(
\begin{array}{ccc}
\lambda_{11} & \lambda_{12} & 0 \\
\lambda_{12} & \lambda_{22} & \lambda_{23} \\
0 & \lambda_{23} & \lambda_{33}
\end{array}
\right) 
{\bf R}_{13} {\bf R}_{23} {\bf R}_{12} 
=
{\bf R}_{13} {\bf R}_{23} {\bf R}_{12}
\left(
\begin{array}{ccc}
\lambda_1 & 0 & 0 \\
0 & \lambda_2 & 0 \\
0 & 0 & \lambda_3
\end{array}
\right)
\ee
from which we can read off the rotation angles
\ba
{s_{13} \over c_{13}} & = & { {\lambda_{23}\lambda_{12}} 
\over {(\lambda_3-\lambda_{22})(\lambda_3-\lambda_{11})-\lambda^2_{12}}}
\nonumber\\
{s_{23} \over c_{23}} & = & { {\lambda_{23}c_{13}+\lambda_{12}s_{13}} 
\over {\lambda_3-\lambda_{22}} }
\nonumber\\
{s_{12} \over c_{12}} & = & { {\lambda_{12}c_{23}
+s_{23}s_{13}(\lambda_{2}-\lambda_{11}) } 
\over {(\lambda_2-\lambda_{11})c_{13}}} \, .
\ea

\end{document}